\documentclass[aps,pre,reprint,showpacs,dvipdfmx,11pt]{revtex4-1}
\usepackage{graphicx}
\usepackage{dcolumn}
\usepackage{amsmath}
\usepackage{amssymb}
\usepackage{color}
\usepackage{subfigure}
\usepackage{bm}
\newcommand{\ave}[1]{\ensuremath{\left\langle#1\right\rangle} }

\newcommand{\const}{\rm const}

\begin{document}
\preprint{HEP/123-qed}
\title{Microscopic determination of macroscopic boundary conditions in Newtonian liquids}
\author{Hiroyoshi Nakano}
\affiliation{Department of Physics, Kyoto University,
Kyoto 606-8502, Japan}
\author{Shin-ichi Sasa}
\affiliation{Department of Physics, Kyoto University,
Kyoto 606-8502, Japan}
\date{\today}

\begin{abstract}
We study boundary conditions applied to the macroscopic dynamics of Newtonian liquids from the view of microscopic particle systems. We assume the existence of microscopic boundary conditions that are uniquely determined from a microscopic description of the fluid and the wall. By using molecular dynamical simulations, we examine a possible form of the microscopic boundary conditions. In the macroscopic limit, we may introduce a scaled velocity field by ignoring the higher order terms in the velocity field that is calculated from the microscopic boundary condition and standard fluid mechanics. We define macroscopic boundary conditions as the boundary conditions that are imposed on the scaled velocity field. The macroscopic boundary conditions contain a few phenomenological parameters for an amount of slip, which are related to a functional form of the given microscopic boundary condition. By considering two macroscopic limits of the non-equilibrium steady state, we propose two different frameworks for determining macroscopic boundary conditions.
\end{abstract}

\pacs{83.50.Rp, 47.10.-g, 05.20.Jj}

\maketitle
\onecolumngrid
\section{Introduction} 
Over the past two decades boundary conditions on solid surfaces have been a focus of study in the field of fluid dynamics ~\cite{neto2005boundary,lauga2007microfluidics,cao2009molecular,bocquet2010nanofluidics}. This focus stems from the remarkable developments of experimental techniques for nano- and micro- scale systems showing the breakdown of the stick boundary condition, specifically, that a fluid at a solid surface has no velocity relative to it~\cite{vinogradova1999slippage,pit2000direct,craig2001shear,tretheway2002apparent,baudry2001experimental,zhu2001rate,zhu2002limits,bonaccurso2002hydrodynamic,cottin2002nanorheology,cheng2002fluid,granick2003slippery,neto2003evidence,vinogradova2003dynamic,lumma2003flow,choi2003apparent,cho2004dipole}. Even Newtonian liquids slip on a solid surface and the boundary condition is far more complicated than conventionally thought. From improvements in experimental techniques and developments in molecular dynamical simulations, many possible boundary conditions for Newtonian liquids have been discovered~\cite{cottin2005boundary,joseph2005direct,huang2006direct,huang2007direct,honig2007no,maali2008measurement,ulmanella2008molecular,steinberger2008nanoscale,cottin2008nanohydrodynamics,vinogradova2009direct,gupta1997shear,barrat1999large,cieplak2001boundary,bocquet1994hydrodynamic,barrat1999influence,cottin2003low,thompson1997general,priezjev2004molecular,martini2008slip,priezjev2009shear,priezjev2010relationship}. The question ``What is the most appropriate boundary condition of Newtonian liquids at solid surfaces?" has attracted a great deal of attention because of its fundamental physical interests and practicality in small-scale fluid dynamics.
However, there are only a few attempts at studying the boundary condition from the perspective of microscopic physical laws.  When we consider the next application of these experimental and numerical results, it is important to give a microscopic foundation of the boundary condition and comprehensively organize these results.

Since the 19th century, the possibility of the breakdown of the stick boundary condition has been discussed. At the center of this discussion, the partial slip boundary condition and the slip length were introduced by Navier~\cite{navier1823}. In the partial slip boundary condition the slip velocity of the fluid at the wall $v_s$ is linearly proportional to the shear rate at the wall $\dot{\gamma}$ as~\cite{lamb1993hydrodynamics,happel2012low,vinogradova1995drainage}
\begin{eqnarray}
v_s = b \dot{\gamma} ,
\end{eqnarray}
where the proportionality constant $b$ is the slip length. The slip length represents the distance at which the fluid velocity extrapolates to zero beyond the surface of the wall. In Navier's partial slip boundary condition, it is assumed that the slip length does not depend on the shear rate~\cite{navier1823,maxwell1879vii}.  By the mid-20th century, the slip length had not been experimentally confirmed and the stick boundary condition had been applied successfully to quantitatively explain numerous macroscopic experiments~\cite{lamb1993hydrodynamics,landau1959course}. However, in the 21st century, sensitive and sophisticated numerical simulations and laboratory experiments of Newtonian liquids in confined geometries have revealed the existence of the slip length and, as a result, the Navier's partial slip boundary condition has been recognized as a more appropriate and practical boundary condition~\cite{bocquet2010nanofluidics,cottin2005boundary,honig2007no,maali2008measurement,cottin2008nanohydrodynamics,vinogradova2009direct}. Much effort has been devoted to the investigation of factors affecting the slip length such as surface roughness~\cite{cottin2003low,wang2003flow,jabbarzadeh2000effect,ponomarev2003surface,priezjev2006influence,priezjev2007effect,niavarani2010modeling} and wettability~\cite{bocquet2010nanofluidics,huang2008water,voronov2008review}.

Further intensive research have discovered the shear dependence of the slip length. The shear-rate-dependent slip was initially intimated in computer simulations at high shear rates~\cite{thompson1997general,priezjev2004molecular,martini2008slip,priezjev2009shear,priezjev2010relationship} and was reported in laboratory experiments~\cite{choi2003apparent,huang2006direct,huang2007direct,ulmanella2008molecular}. These studies indicate that the slip length is independent of the shear rate only when the shear rate is small enough~\cite{bocquet1994hydrodynamic}.

Based on these achievements, research on boundary conditions is expected to move to a new stage. The shear dependence of the slip length is obviously a breakdown of Navier's partial slip boundary condition. As advances in experimental techniques replaced the stick boundary condition with Navier's partial slip boundary condition as the fundamental boundary condition, more advanced experimental technique will replace Navier's partial slip boundary condition with a more fundamental boundary condition. At this time, the microscopic foundation of boundary conditions is of practical importance. Thus, the first problem we are tasked with is ``to determine the microscopic boundary condition from the viewpoint of microscopic particle systems."

Here, even if we obtain a microscopic boundary condition, the boundary conditions we conventionally used for a macroscopic description are still worthwhile. Such macroscopic boundary conditions have been applied to obtain satisfactory results from the macroscopic point of view in many situations. Therefore, whenever we impose an extent of the measurement accuracy from the macroscopic point of view, the system can be characterized by the macroscopic boundary condition rather than the microscopic boundary condition. We should define this measurement accuracy as a mathematical concept so that we can connect the macroscopic boundary condition with the microscopic boundary condition. Thus, the second problem we tackle is “to derive the macroscopic boundary conditions from the microscopic boundary condition by formulating proper macroscopic limits."

In this paper, we propose a framework to organize the macroscopic boundary conditions for a simple case, specifically, uniform shear flow. The starting point should be the microscopic boundary condition. Since it is unknown, we first introduce a tentative fundamental boundary condition that is consistent with results obtained previously in numerical simulations and laboratory experiments. For this purpose, we use molecular dynamical simulation. Then we introduce the measurement accuracy in the uniform shear flow as a mathematical concept. By using this framework, we discuss what kind of boundary conditions should be used in a given situation.

The key idea is to introduce a relation between the measurement accuracy and the system-size-dependence of the velocity fields in the infinite volume limit of the uniform shear flow. By ignoring the higher terms of the velocity fields in the system size, we formulate the measurement accuracy. Then we can obtain the macroscopic boundary condition that satisfies the required measurement accuracy. We notice that the macroscopic boundary conditions depend on the choice of the infinite volume limit of the uniform shear flow and the order of terms to be left. We develop two different frameworks of the macroscopic boundary conditions by considering two different infinite volume limits of the uniform shear flow.

The remainder of this paper is organized as follows. In Sec.~\ref{sec:Setup and question}, the setup of our model is introduced. We explain the problems to be studied in this paper in terms of our setup. In Sec.~\ref{sec:Molecular dynamical simulation}, we describe the determination of the microscopic boundary condition by using the molecular dynamical simulations. In Sec.~\ref{sec:Macroscopic boundary condition}, we determine the macroscopic boundary conditions based on the microscopic boundary condition. Secs.~\ref{sec:summary} and \ref{sec:discussion} are devoted to a brief summary and discussion.

\section{Setup and question}
\label{sec:Setup and question}

\subsection{Model}
\begin{figure}
\centering
\includegraphics[width=0.7\linewidth]{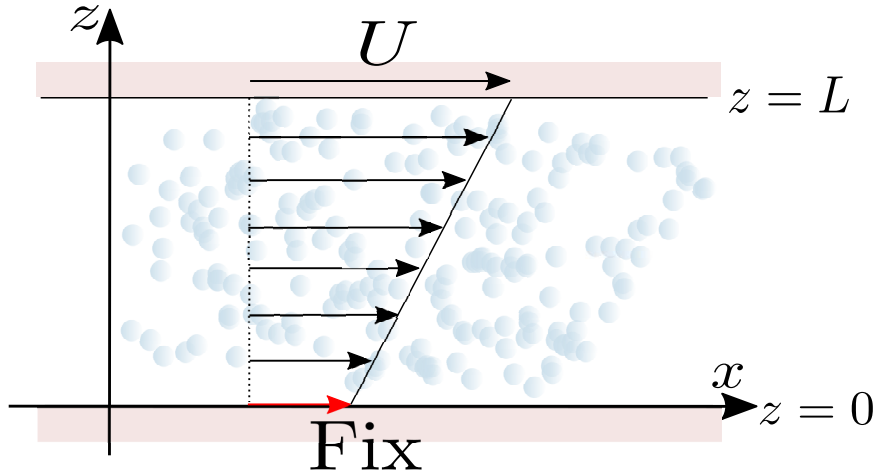}
\caption{Schematic illustration of our model. }
\label{fig1}
\end{figure}

We introduce a model for studying boundary conditions for fluid dynamics.  A schematic illustration is shown in Fig.~\ref{fig1}. 
The fluid consists of $N$ particles that are confined to an $L_x\times L_y\times L_z$ cubic box. We impose periodic boundary conditions along the $x$ and $y$ directions and introduce two parallel walls so as to confine particles in the $z$ direction. We represent the two walls as potential forces acting on the particles. Let $(\bm{r}_i,\bm{p}_i)$, $(i=1,2,\cdots,N)$, be the position and momentum of the $i$th particle. The Hamiltonian of the system is given by
\begin{eqnarray}
H = \sum_{i=1}^{N} \frac{\bm{p}_i^2}{2m} + U((\bm{r}_i)_{i=1}^N)
\end{eqnarray}
with
\begin{eqnarray}
U((\bm{r}_i)_{i=1}^N) &\equiv& \sum_{i<j} V_{\rm FF}(|\bm{r}_i-\bm{r}_j|) + \sum_{i=1}^N U_{\rm BW}(\bm{r}_i) + \sum_{i=1}^N U_{\rm TW}(\bm{r}_i) .
\label{eq:potential all}
\end{eqnarray}
$V_{\rm FF}(r)$ describes an interaction potential between two particles. $U_{\rm BW}(\bm{r})$ and $U_{\rm TW}(\bm{r})$ represent a $z=0$ wall potential and a $z=L_z$ wall potential, respectively. In region B near the $z=L_z$ wall, which is given by $[0,L_x]\times[0,L_y]\times[L,L_z]$, we apply the Langevin thermostat and the external force $f$ along the $x$-axis. We assume that $U_{\rm TW}(\bm{r})$ has non-zero value only in region B. Then, the particles obey the Langevin equation
\begin{eqnarray}
m\frac{d^2 r^{\alpha}_i}{dt^2} = - \sum_{j(\neq i)} \frac{\partial V_{\rm FF}(|\bm{r}_i-\bm{r}_j|)}{\partial r^{\alpha}_i} - \sum_{i=1}^N \frac{\partial U_{\rm BW}(\bm{r}_i)}{\partial r^{\alpha}_i}
\end{eqnarray}
for $z \in [0,L]$, and
\begin{eqnarray}
m\frac{d^2 r^{\alpha}_i}{dt^2} &=& - \sum_{j(\neq i)} \frac{\partial V_{\rm FF}(|\bm{r}_i-\bm{r}_j|)}{\partial r^{\alpha}_i} - \sum_{i=1}^N \frac{\partial U_{\rm TW}(\bm{r}_i)}{\partial r^{\alpha}_i} + f^{\alpha} - \zeta \frac{d r^{\alpha}_i}{dt} + \xi^{\alpha}_i(t) 
\label{eq: dynamics x 0.9L<}
\end{eqnarray}
for $z \in [L,L_z]$, where $\bm{f}=(f,0,0)$, $\bm{\xi}_i$ represents thermal noise satisfying
\begin{eqnarray}
\langle \xi_i^{\alpha}(t) \xi_j^{\beta}(t') \rangle = 2 \zeta k_{\rm B} T \delta_{ij} \delta^{\alpha \beta} \delta(t-t'),
\end{eqnarray}
where $k_{\rm B}$ is the Boltzmann constant, $T$ the temperature of the thermostat, and $\zeta$ the friction coefficient.

\subsection{observed quantity}
We concentrate on the velocity vector field and stress tensor field in the steady state. This subsection summarizes the definition of these quantities. Let $\hat{\rho}(\bm{r};\Gamma_t)$, $\hat{\pi}^a(\bm{r};\Gamma_t)$, and $\hat{J}^{ab}(\bm{r};\Gamma_t)$ denote the microscopic mass density, momentum density and momentum current density at a given point $\bm{r}$, respectively, for a given microscopic configuration $\Gamma_t \equiv (\bm{r}_1(t),\cdots,\bm{r}_N(t),\bm{p}_1(t),\cdots,\bm{p}_N(t))$ at time $t$; see Appendix~\ref{sec:expression of microscopic density} for details of these definitions. We consider the temporal and spatial average of these microscopic fields. In particular, we consider the $z$-dependence of the averaged local quantities. We perform spatial average in the slab with bin width $\Delta z$ at the center $z$ and temporal average for a time interval $\tau$ in the steady state. For example, the averaged mass density at any $z$ is given by
\begin{eqnarray}
\rho(z) &=& \ave{\hat{\rho}(z)}_{\rm ss} = \frac{1}{\tau} \int_{0}^{\tau} dt \frac{1}{L_xL_y} \int_0^{L_x} dx \int_0^{L_y}dy \frac{1}{\Delta z} \int_{z-\Delta z/2}^{z+\Delta z/2} dz \hat{\rho}(\bm{r};\Gamma_t),
\label{eq:example of rho}
\end{eqnarray}
where the system is assumed to be in steady state at $t=0$. Similarly, we give the averaged momentum density $\pi^a(z)$ and momentum current $J^{ab}(z)$. Then, we define velocity $v^a(z)$ and stress $\sigma^{ab}(z)$ at $z$ as
\begin{eqnarray}
v^a(z) = \frac{\pi^a(z)}{\rho(z)},
\end{eqnarray}
\begin{eqnarray}
\sigma^{ab}(z) = -J^{ab}(z) + \rho(z) v^a(z) v^b(z).
\end{eqnarray}
We assume that the velocity field is parallel to the $x$-direction sufficiently far away from the wall. We focus on $v^x(z)$ and $\sigma^{xz}(z)$.

\subsection{Problem}
\label{sec:Question}
This subsection explains the problem to be studied in the remainder of this paper in terms of the quantities defined above.

We refer to the region sufficiently far from the walls as the bulk. Our chief concern is the velocity and stress profiles $(v^x(z),\sigma^{xz}(z))$ of the bulk in the steady state. Fluid mechanics is the theory for describing the macroscopic behaviors of these quantities. For our setup, the constitutive equation is given by
\begin{eqnarray}
\sigma^{xz} = \eta \frac{d v^x}{dz}
\label{eq:Constitutive equation in bulk}
\end{eqnarray}
except for a region near the walls, where $\eta$ is a dynamical viscous coefficient. Then, we extrapolate the velocity field in the bulk to the whole region $[0,L_x]\times[0,L_y]\times[0,L]$ while retaining the relation (\ref{eq:Constitutive equation in bulk}).  Let the extrapolated velocity at $z=L$ be given by
\begin{eqnarray}
v^x(L) = U.
\label{eq:boundary condition for upper wall}
\end{eqnarray}
Since we obtain any $U$ by controlling the external force $f$ in our setup, we may treat $U$ as a parameter. Then, we focus on the boundary condition at the $z=0$ wall. Because forces are balanced in the steady state, the shear stress is independent of the $z$-coordinate:
\begin{eqnarray}
\sigma^{xz}(z) = \sigma^{xz} = {\rm const}.
\label{eq:balance of forces}
\end{eqnarray}
From (\ref{eq:Constitutive equation in bulk}), (\ref{eq:boundary condition for upper wall}) and (\ref{eq:balance of forces}), we characterize the extrapolated velocity field by
\begin{eqnarray}
v^x(z) = \frac{\sigma^{xz}}{\eta} (z-L) + U,
\label{eq:velocity field containing the unknown constant}
\end{eqnarray}
where $\eta$ is assumed to be known. When we observe stress $\sigma^{xz}$, we obtain the extrapolated velocity field by using (\ref{eq:velocity field containing the unknown constant}) with $\eta$ and $U$. Thus, if a boundary condition determines the extrapolated velocity field $v^x(z)$, the boundary condition should be related to $\sigma^{xz}$, which we express as $\sigma^{xz}(U)$ with $\eta$ fixed. We emphasize that we study the extrapolated velocity field instead of the real velocity field, because our main concern is $(v^x(z),\sigma^{xz}(z))$ in the bulk, and not the real velocity field near the walls. Hereafter, for simplicity, we refer to the extrapolated velocity field as the velocity field.

We remark that the boundary condition may depend on the measurement accuracy or the scale of interest. For example, there is a case that finite $|v^x(0)|$ cannot be observed for a given accuracy in an investigation for a phenomenon. $\sigma^{xz}(U)$ should be determined in accordance with the required accuracy of $v^x(z)$. We refer to such boundary conditions as the macroscopic boundary condition. Moreover, it is reasonable to conjecture that there is a boundary condition determined only by the microscopic setup, independent of the scale of interest. If we demand greater accuracy in $v^x(z)$, then we should use this microscopic boundary condition. In this paper, we explore the most appropriate microscopic boundary condition and study the macroscopic boundary condition based on the appropriate condition.

\section{Molecular dynamical simulation}
\label{sec:Molecular dynamical simulation}
\subsection{Preliminaries}
We perform numerical simulations with the following potentials in (\ref{eq:potential all}). First, the interaction between two particles $V_{\rm FF}(r)$ is given by the truncated Lennard-Jones potential with a cut-off length $r_c$:
\begin{eqnarray}
V_{\rm FF}(r) \equiv 4 \epsilon_{} \bigg(\Big(\frac{\sigma_{}}{r}\Big)^{12} -c_{} \Big(\frac{\sigma_{}}{r}\Big)^6 + C^{(2)}_{\rm FF} r^2 + C^{(0)}_{\rm FF} \bigg)
\label{eq:truncated Lennard-Jones potential}
\end{eqnarray}
for $r<r_c$, and $V_{\rm FF}(r)=0$ otherwise. $C^{(2)}_{FF}$ and $C^{(0)}_{FF}$ are determined by the condition $V_{\rm FF}(r_c)=0$ and $V_{\rm FF}'(r_c)=0$~\cite{watanabe2011efficient}. Second, the $z=0$ wall consists of $N_w$ material points, which are fixed on the square lattice in the $z=0$ plane. The lattice constant is denoted by $a$. Let $\bm{q}_i$ $(i=1,2,\cdots,N_w)$ be the position of the material points. The interaction potential between a material point and a fluid particle $V_{\rm BW}(r)$ is given by the same form as (\ref{eq:truncated Lennard-Jones potential}) with $\epsilon_{\rm BW}$, $\sigma_{\rm BW}$ and $c_{\rm BW}$. Then, $ U_{\rm BW}(\bm{r})$ is expressed by
\begin{eqnarray}
U_{\rm BW}(\bm{r}) \equiv \sum_{j=1}^{N_w} V_{\rm BW}(|\bm{r}-\bm{q}_j|),
\label{eq:UBW}
\end{eqnarray}
where $\sigma_{\rm BW}$ is given by
\begin{eqnarray}
\sigma_{\rm BW} \equiv \frac{a+\sigma_{}}{2}
\end{eqnarray}
so that the lattice constant $a$ is treated as the diameter of the particles constituting the $z=0$ wall. Finally, the potential between the $z=L_z$ wall and a fluid particle is given by
\begin{eqnarray}
U_{\rm TW}(\bm{r}) &=& 4 \epsilon_{\rm TW} \bigg(\Big(\frac{\sigma_{\rm TW}}{L_z-z}\Big)^{12} - \Big(\frac{\sigma_{\rm TW}}{L_z-z}\Big)^6 + C^{(2)}_{\rm TW} (L_z-z)^2 + C^{(0)}_{\rm TW} \bigg)
\end{eqnarray}
for $z>L_z-r_c$; otherwise, $U_{\rm TW}(\bm{r})=0$.

In numerical simulations, all the quantities are converted to dimensionless forms by setting $m=\sigma_{}=\epsilon_{}=1$. We fix $L_x=30.0\sigma_{}$, $L_y=30.0\sigma_{}$, $L_z=24.0\sigma_{}$, and $L=20.0 \sigma_{}$. The particle number is set to $N=16200$, which corresponds to particle number density $\rho = 0.75 \sigma_{}^{-3}$. The temperature and the friction coefficient of the Langevin thermostat are set to $k_{\rm B} T/\epsilon_{}=1.1$ and $\zeta = 1.0 \sqrt{\epsilon m}/\sigma$, respectively. The potential parameters are fixed to $c_{}=1.0$, $\epsilon_{\rm TW}/\epsilon_{}=\epsilon_{\rm BW}/\epsilon_{}=0.6$, and $\sigma_{\rm TW}/\sigma=1.0 $. The cutoff distance is set to $r_c=2.5\sigma_{}$. Then, we characterize the $z=0$ wall by the value of $a$ and $c_{\rm BW}$.

\subsection{Microscopic boundary condition}
\label{sec:Microscopic boundary condition}
\begin{figure}
\centering
\includegraphics[width=0.7\linewidth]{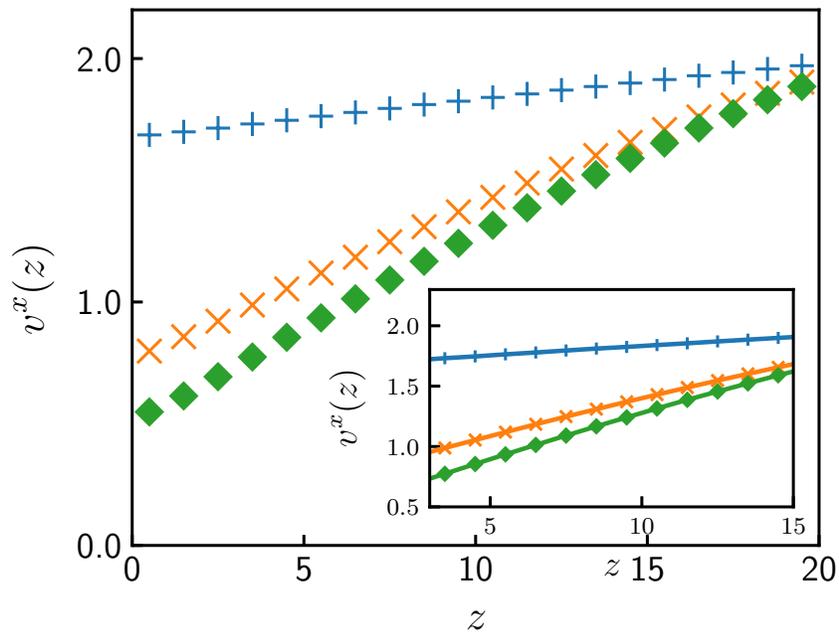}
\caption{Velocity profiles for the applied force $f=2.0$. The wall parameters are chosen as $(a,c_{\rm BW}) = (0.5,0.6)$ (blue), $(0.6,1.0)$ (orange) and $(0.7,1.0)$ (green). Inset: linear fits of the velocity profile away from the walls.}
\label{fig:uniform shear flow}
\end{figure}

\begin{figure}
\centering
\includegraphics[width=0.7\linewidth]{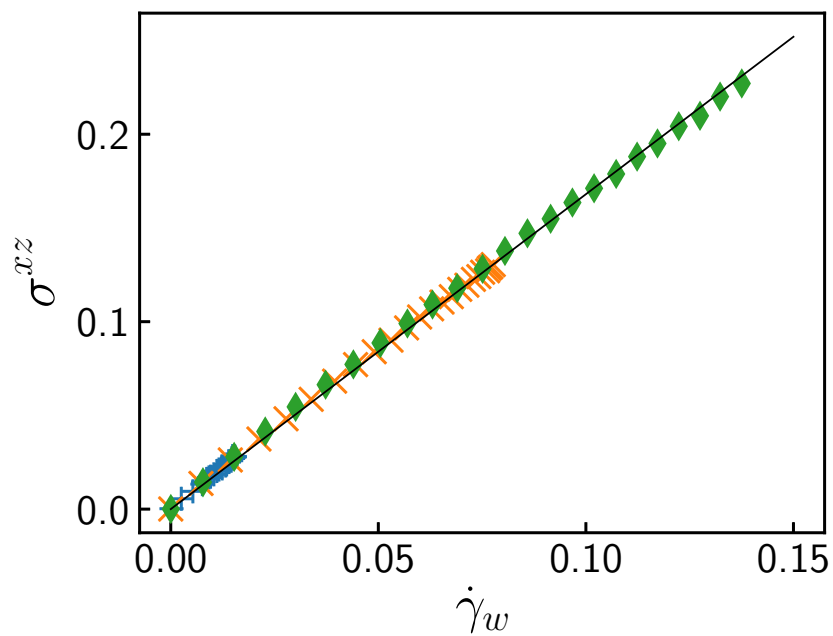}
\caption{Stress tensor as a function of shear rate. The applied force $f$ is varied from $0.2$ to $4.8$. The wall parameters are chosen as $(a,c_{\rm BW}) = (0.5,0.6)$ (blue), $(0.6,1.0)$ (orange) and $(0.7,1.0)$ (green).}
\label{fig:constitutive equation in bulk}
\end{figure}

We study the behavior near the $z=0$ wall.
Figure~\ref{fig:uniform shear flow} shows examples of the velocity profile in the steady state, with $f=2.0$ and $(a,c_{\rm BW}) = (0.5,0.6),\ (0.6,1.0), \ {\rm and}\  (0.7,1.0)$. 
The velocity profiles in $3\leq z \leq 15$ (inset of Fig.~\ref{fig:uniform shear flow}) are well fitted linearly. This suggests that uniform shear flow appears in the region $3\leq z \leq 15$. Therefore, we identify this region with the bulk. Figure~\ref{fig:constitutive equation in bulk} shows the shear stress as a function of shear rate in the bulk. From Fig.~\ref{fig:constitutive equation in bulk}, we find that (\ref{eq:Constitutive equation in bulk}) holds and $\eta$ is independent of wall parameters. We note that $\eta$ is independent of the shear rate in the shear rate range used in this paper.

\begin{figure}
\centering
\includegraphics[width=0.7\linewidth]{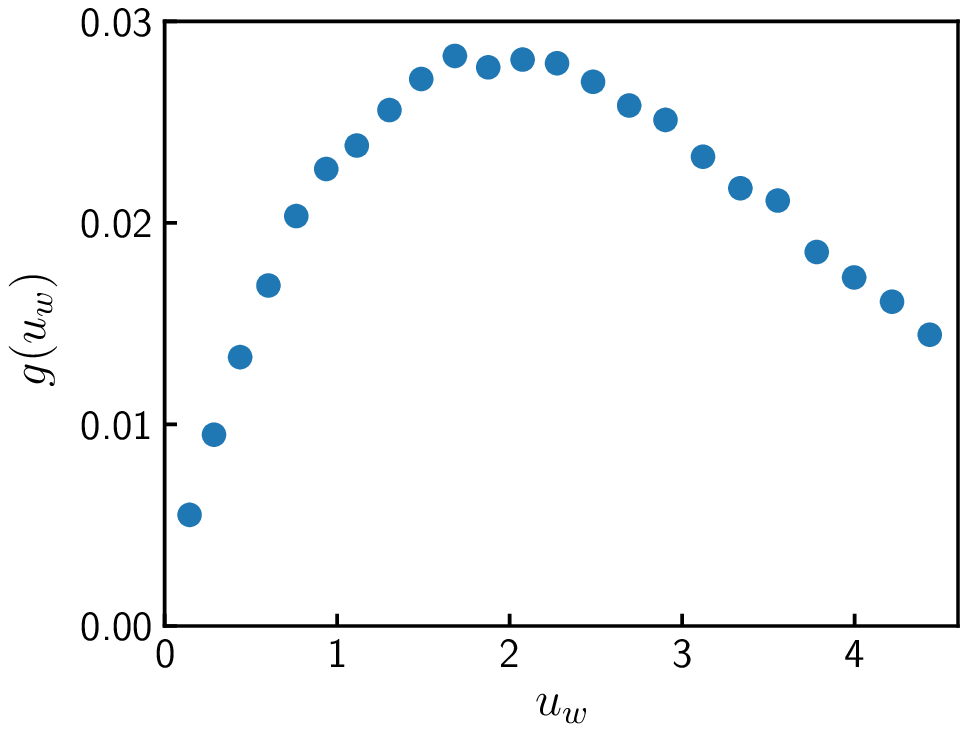}
\caption{Plot of $g(u_w)$. The wall parameters and the applied force are chosen as $a=0.5,c_{\rm BW}=0.6$ and $0.2\leq f\leq 4.8$, which is the same as Fig.~\ref{fig:constitutive equation in bulk}.}
\label{fig:uw vs gamma}
\end{figure}

We consider the boundary condition at $z=0$ that is consistent with the velocity profiles measured above. The observation in Sec.~\ref{sec:Question} suggests that the microscopic boundary condition is expressed in terms of the shear stress $\sigma^{xz}$ as a function of the fluid velocity. By noting that the boundary condition is expected to be locally given, we find that the simplest boundary condition is given by
\begin{eqnarray}
\sigma^{xz} = g(u_w),
\label{eq:constitutive equation in boundary region}
\end{eqnarray}
where $u_w$ is the slip velocity extracted from the extrapolated velocity field. In Fig.~\ref{fig:uw vs gamma}, we plot $g(u_w)$ for the wall with $(a,c_{\rm BW})=(0.5,0.6)$ as $f$ increases from $0.2$ to $4.8$.
We next show that the velocity field $v^x(z)$ is uniquely determined when $g(u_w)$ is given, which is a necessary condition for a boundary condition. By combining (\ref{eq:velocity field containing the unknown constant}) with (\ref{eq:constitutive equation in boundary region}), we obtain
\begin{eqnarray}
\eta \frac{U-u_w}{L} = g(u_w).
\label{eq:determination of uw}
\end{eqnarray}
By solving (\ref{eq:determination of uw}), we obtain $u_w$. Given $u_w$, $v^x(z)$ is written as
\begin{eqnarray}
v^x(z) = \frac{U-u_w}{L} z + u_w.
\label{eq:vxz most useful form}
\end{eqnarray}
Therefore, we interpret (\ref{eq:constitutive equation in boundary region}) to be a microscopic boundary condition with $g(u_w)$, the functional form of which is specific to details of the wall and particles.

We remark on some equivalent expressions of (\ref{eq:constitutive equation in boundary region}). We first note that the previous studies~\cite{bocquet1994hydrodynamic,navier1823,lamb1993hydrodynamics,happel2012low,vinogradova1995drainage} proposed a boundary condition
\begin{eqnarray}
v^x \Big|_{z=0} = b \frac{\partial v^x}{\partial z} \Big|_{z=0} ,
\label{eq:partial slip boundary condition in md}
\end{eqnarray}
instead of (\ref{eq:constitutive equation in boundary region}), where $b$ corresponds to the slip length. 
We note that the slip length $b$ may depend on the macroscopic velocity field. By using (\ref{eq:Constitutive equation in bulk}), (\ref{eq:boundary condition for upper wall}), and (\ref{eq:partial slip boundary condition in md}), we find that the velocity field is expressed in terms of the slip length $b$ as
\begin{eqnarray}
v^x(z) = \frac{U}{L+b}(z+b) .
\label{eq:velocity field with partial slip boundary condition}
\end{eqnarray}
By fitting the velocity profile $v^x(z)$ measured in numerical simulations to (\ref{eq:velocity field with partial slip boundary condition}), we obtain the slip length $b$.
The slip length as a function of $u_w$, $b(u_w)$, is equivalent to the microscopic boundary condition (\ref{eq:constitutive equation in boundary region}). This is because (\ref{eq:Constitutive equation in bulk}) and (\ref{eq:constitutive equation in boundary region}) lead to (\ref{eq:partial slip boundary condition in md}) with 
\begin{eqnarray}
b(u_w) = \frac{\eta u_w}{g(u_w)}.
\label{eq:relation between b and gamma}
\end{eqnarray}
As other cases, some previous studies considered the slip length $b$ as a function of shear rate near the wall, $\dot{\gamma}_w$~\cite{thompson1997general,priezjev2004molecular,priezjev2007effect}. We rewrite (\ref{eq:partial slip boundary condition in md}) in terms of $\dot{\gamma}_w$ as
\begin{eqnarray}
U - \dot{\gamma}_w L = b(\dot{\gamma}_w) \dot{\gamma}_w.
\label{eq:equation0}
\end{eqnarray}
If $b(\dot{\gamma}_w)$ is given, then we calculate $\dot{\gamma}_w$ as a function of $U$ and $L$ by solving (\ref{eq:equation0}). By recalling $u_w = U - \dot{\gamma}_w L$ and by comparing (\ref{eq:constitutive equation in boundary region}) to (\ref{eq:equation0}), we construct $b(\dot{\gamma}_w)$ from $g(u_w)$ as
\begin{eqnarray}
b(\dot{\gamma}_w) = \frac{h(\eta \dot{\gamma}_w)}{\dot{\gamma}_w}
\label{eq:h g inverse}
\end{eqnarray}
where $h(\sigma^{xz})$ is the multi-valued function that yields possible values of $u_w$ satisfying (\ref{eq:constitutive equation in boundary region}) for a given $\sigma^{xz}$. Equation (\ref{eq:h g inverse}) implies that (\ref{eq:constitutive equation in boundary region}) and (\ref{eq:partial slip boundary condition in md}) with $b(\dot{\gamma}_w)$ are equivalent.

\begin{figure}
\centering
\includegraphics[width=0.9\linewidth]{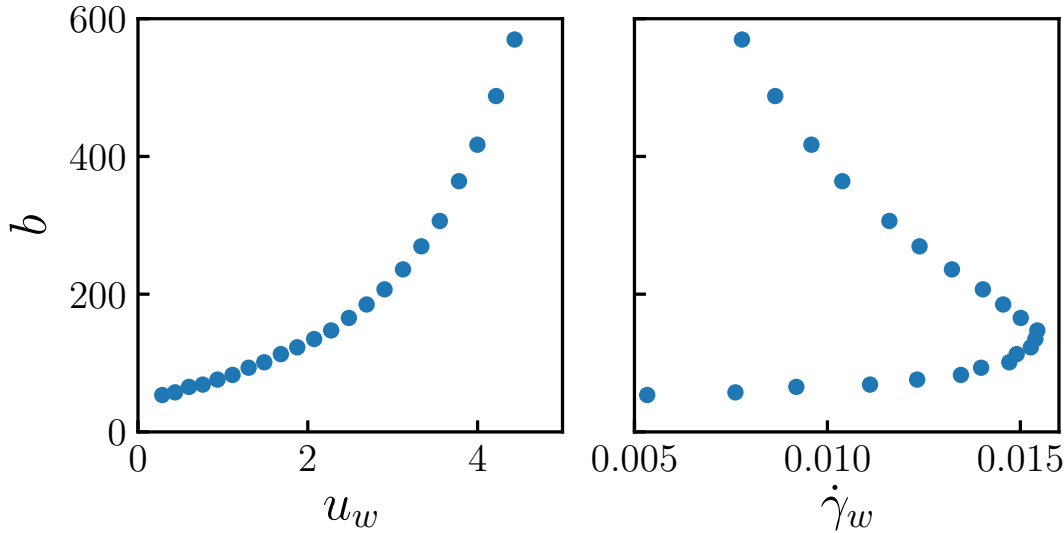}
\caption{Slip length as a function of local velocity at the $z=0$ wall, $u_w$, (left-hand side) and local shear rate at the $z=0$ wall, $\dot{\gamma}_w$ (right-hand side). The parameter settings are the same as those in Fig.~\ref{fig:uw vs gamma}.}
\label{fig:Slip Length}
\end{figure}

In Fig.~\ref{fig:Slip Length}, we plot $b$ as a function of $u_w$ and $\dot{\gamma}_w$ for the same parameters as Fig.~\ref{fig:uw vs gamma}.
From Figs.~\ref{fig:uw vs gamma} and \ref{fig:Slip Length}, we find that the microscopic boundary condition exhibits a non-linear behavior. Specifically, Fig.~\ref{fig:Slip Length} indicates that the slip length depends non-linearly on $u_w$ or $\dot{\gamma}_w$ and reaches a value more than ten times the system size. This behavior is consistent with some experimental results~\cite{neto2005boundary,lauga2007microfluidics,bocquet2010nanofluidics,cao2009molecular}. We remark that the previous numerical simulations~\cite{thompson1997general,priezjev2004molecular,priezjev2009shear} found a critical shear rate $\dot{\gamma}_c$ at which $b(\dot{\gamma}_w)$ diverges as $\dot{\gamma}_w \to \dot{\gamma}_c$. We conjecture that the results of our simulation are consistent with that of these studies. From the right-hand side of Fig.~\ref{fig:Slip Length}, we find that $d b(\dot{\gamma}_w)/d\dot{\gamma}_w$ diverges. We consider that the divergence of $b(\dot{\gamma}_w)$ reported in the previous studies corresponds to the divergence of $d b(\dot{\gamma}_w)/d\dot{\gamma}_w$ in our simulation. In Appendix.~\ref{sec:scaling law at high shear rates}, we demonstrate the correspondence between our simulation and the previous studies by focusing on the scaling law as $\dot{\gamma}_w \to \dot{\gamma}_c$ reported in some studies~\cite{thompson1997general,priezjev2004molecular,priezjev2009shear}.

In Sec.~\ref{sec:Macroscopic boundary condition : hydrodynamic limit}, we shall focus on the non-linear behavior of the microscopic boundary condition, particularly on the existence of the maximum of $g(u_w)$. The point of $b(u_w)$ that corresponds to the maximum point of $g(u_w)$ is calculated from (\ref{eq:relation between b and gamma}). We find that this point of $g(u_w)$ has a simple graphical interpretation in contrast to that of $b(u_w)$ (see Fig.~\ref{fig:uw vs gamma} and the left-hand side of Fig.~\ref{fig:Slip Length}). Also, the corresponding point in $b(\dot{\gamma}_w)$ is the point that the first derivative in $\dot{\gamma}_w$, $d b(\dot{\gamma}_w)/d\dot{\gamma}_w$, diverges (see Appendix.~\ref{sec:scaling law at high shear rates}). The divergence of $d b(\dot{\gamma}_w)/d\dot{\gamma}_w$ provides this point in $b(\dot{\gamma}_w)$ with a simple graphical interpretation. Therefore, we expect that $g(u_w)$ or $b(\dot{\gamma}_w)$ is more useful than $b(u_w)$ for the discussion using graphs. Furthermore, using $b(\dot{\gamma}_w)$ is more mathematically inconvenient than $g(u_w)$ because $b(\dot{\gamma}_w)$ is a two-valued function in $\dot{\gamma}_w$. Therefore, in the reminder of this paper, we use (\ref{eq:constitutive equation in boundary region}) with given $g(u_w)$ as the microscopic boundary condition.

\section{Macroscopic boundary condition}
\label{sec:Macroscopic boundary condition}
The microscopic boundary condition is uniquely determined from the microscopic description of the fluid and the wall. That is, $g(u_w)$ is uniquely determined from a given microscopic model. As we change the scale of interest from the microscopic to the macroscopic, we may use the macroscopic boundary condition instead of the microscopic boundary condition. In this section, we study how the macroscopic boundary condition appears, depending on the choice of the scale of interest. For this, we introduce how to choose the scale of interest as a mathematical concept.

\subsection{Choice of the scale of interest}
We focus on the $L$-dependence of the velocity field $\bar{v}^x(\bar{z})$ as a function of $\bar{z} \equiv z/L$:
\begin{eqnarray}
\bar{v}^x(\bar{z}) = (U-u_w) \bar{z} + u_w,
\label{eq:bar vx}
\end{eqnarray}
where we have used (\ref{eq:vxz most useful form}). We introduce the scaled velocity field by ignoring higher order terms of $\bar{v}^x(\bar{z})$ in $L$ depending on the scale of interest. The macroscopic boundary condition is determined so that the scaled velocity field is obtained in the standard fluid dynamics. We notice that the macroscopic boundary condition depends on the choice of the terms retained in the scaled velocity field. From (\ref{eq:bar vx}), we find that the $L$-dependence of $\bar{v}^x(\bar{z})$ is determined from that of $U$ and $u_w$. This implies that the macroscopic boundary condition is related to the $L$-dependence of $u_w$. As described in Sec.~\ref{sec:Microscopic boundary condition}, we obtain $u_w$ by solving (\ref{eq:determination of uw}). Because the $L$-dependence of $u_w$ is connected to the functional form of $g(u_w)$ through (\ref{eq:determination of uw}), we can define the macroscopic boundary condition by the $L$-dependence of $u_w$ or the functional form of $g(u_w)$.

In the remainder of this section, we consider two macroscopic limits in the non-equilibrium steady state that is subjected to the uniform shear flow. In each macroscopic limit, we study the macroscopic boundary condition.

\begin{figure*}[htbp]
\begin{center}
\subfigure[]{
\includegraphics[width=0.45\linewidth]{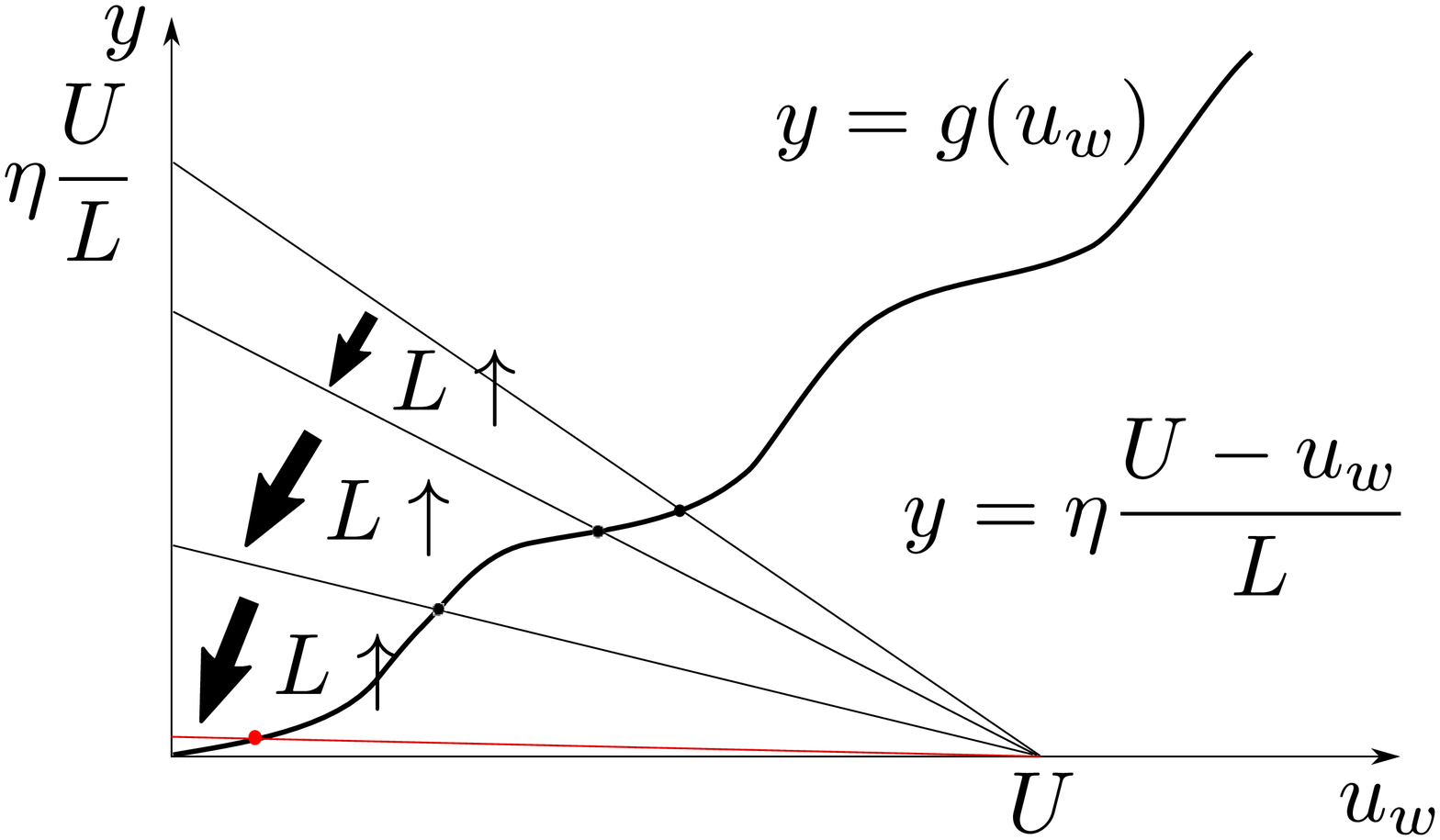} 
\label{fig:type 1-3}}
\subfigure[]{
\includegraphics[width=0.48\linewidth]{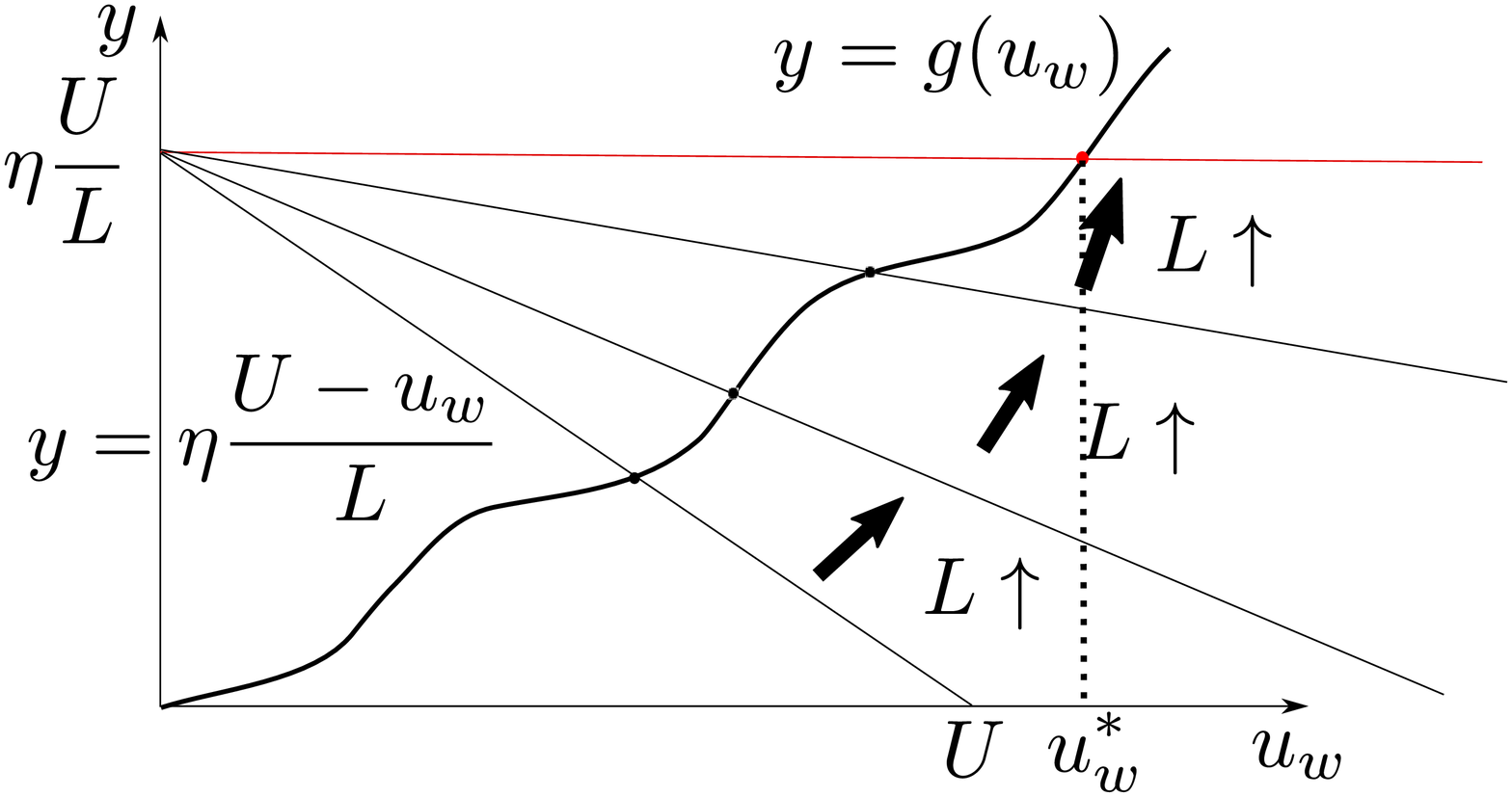}
\label{fig:type 1-2}}
\end{center}
\caption{Schematic graph of $y=g(u_w)$ and $y=\eta (U-u_w)/L$. The intersection of these graphs corresponds to the solution of (\ref{eq:determination of uw}): (a) behavior of the solution in the quasi-equilibrium limit; (b) behavior of the solution in the hydrodynamic limit.}
\label{fig:type 1-2 and 1-3}
\end{figure*}

\subsection{Macroscopic boundary condition: quasi-equilibrium limit}
\label{sec:Macroscopic boundary condition : quasi-equilibrium limit}
The first macroscopic limit is the quasi-equilibrium limit:
\begin{eqnarray}
L \to \infty, \ U = \const, \ \rho = \const.
\label{eq:infinite volume limit A}
\end{eqnarray}
We focus on the $O(L^{-1})$ terms of the velocity fields $\bar{v}^x(\bar{z})$ as the scale of interest. In this section, $\simeq$ indicates equality up to $o(L^{-1})$ terms. 

We define three boundary conditions by noting the $L$-dependence of $u_w$ in the quasi-equilibrium limit (\ref{eq:infinite volume limit A}): stick boundary condition $u_w = o(L^{-1})$, partial slip boundary condition $u_w = O(L^{-1})$, and perfect slip boundary condition $L u_w \to \infty$. Then, the stick boundary condition $u_w = o(L^{-1})$ implies 
\begin{eqnarray}
\bar{v}^x(\bar{z}) \simeq U \bar{z},
\end{eqnarray}
which is consistent with the standard stick boundary condition in hydrodynamics.

We consider a relationship between the $L$-dependence of $u_w$ and the functional form of $g(u_w)$. We focus on the case in which the functional form of $g(u_w)$ is given by Fig.~\ref{fig:type 1-2 and 1-3}. In Fig.~\ref{fig:type 1-3}, we present the two graphs $y=g(u_w)$ and 
\begin{eqnarray}
y=\eta \frac{U-u_w}{L}.
\end{eqnarray}
The intersection of the two graphs corresponds to the solution of (\ref{eq:determination of uw}). Since $U$ is fixed and $\eta U/L$ approaches $0$ in the quasi-equilibrium limit (\ref{eq:infinite volume limit A}), the $L$-dependence of the solution is determined by the behavior of $y=g(u_w)$ near $y=0$ (See Fig.~\ref{fig:type 1-3}). Then, we consider the three cases of the behavior of $y=g(u_w)$ near $y=0$. 

First, let $g(u_w)$ be expanded around $u_w=0$ as  
\begin{eqnarray}
g(u_w) = g_1 u_w + \frac{1}{2} g_2 u_w^2 + \cdots.
\label{eq:expansion of gamma}
\end{eqnarray}
We assume $g(0)=0$ so that the fluid exerts no force on the wall if $u_w=0$. We consider the case $g_1 \neq0$. By substituting (\ref{eq:expansion of gamma}) into (\ref{eq:determination of uw}) and solving for $u_w$, we obtain
\begin{eqnarray}
u_w \simeq \frac{\eta}{g_1} \frac{U}{L}.
\label{eq:uw partial slip}
\end{eqnarray}
As $u_w$ is of order $L^{-1}$, we find that $g(u_w)$ with $g_1 \neq0$ corresponds to the partial slip boundary condition. 

We next consider the case $g_1=0$ and $g_2 \neq 0$. By the similar calculation, we obtain
\begin{eqnarray}
u_w = O(L^{-1/2})
\label{eq:Ldependence of uw L12}
\end{eqnarray}
in the quasi-equilibrium limit (\ref{eq:infinite volume limit A}). Therefore, we find that $g(u_w)$ with $g_1=0$ and $g_2 \neq 0$ corresponds to the perfect slip boundary condition.  

Finally, let the first derivative of $g(u_w)$ diverge at $u_w=0$, as in for instance
\begin{eqnarray}
g(u_w) \simeq u_w^a
\end{eqnarray}
near $u_w=0$, where $0<a<1$. By solving (\ref{eq:determination of uw}), we obtain
\begin{eqnarray}
u_w = O(L^{-1/a})
\end{eqnarray}
in the quasi-equilibrium limit (\ref{eq:infinite volume limit A}), which corresponds to the stick boundary condition. These results indicate that the boundary condition is determined only by the analyticity of $g(u_w)$ in $u_w=0$.

We remark that the relationship between the boundary condition defined above and the slip length. We consider the case satisfying the partial slip boundary condition. Since we need to know the linear term of $g(u_w)$ to obtain (\ref{eq:uw partial slip}), we rewrite (\ref{eq:determination of uw}) as
\begin{eqnarray}
\frac{\partial v^x}{\partial z} \Big|_{z=0} = \frac{1}{b_1} v^x \Big|_{z=0}
\label{eq:traditional partial slip boundary condition b1}
\end{eqnarray}
with 
\begin{eqnarray}
b_1 \equiv \frac{\eta}{g_1}.
\label{eq:b1}
\end{eqnarray}
We keep in mind that $o(L^{-1})$ terms of $\bar{v}^x(\bar{z})$ are irrelevant for the solutions of the Navier--Stokes equation with boundary condition (\ref{eq:traditional partial slip boundary condition b1}). That is, although the form of the boundary condition (\ref{eq:traditional partial slip boundary condition b1}) is the same as Navier's partial slip boundary condition, i.e., constant slip length, these boundary conditions are different in whether we impose an extent of $\bar{v}^x(\bar{z})$ to be focused. Similarly, we may rewrite the perfect slip boundary condition in the form (\ref{eq:traditional partial slip boundary condition b1}). For example, for $g_1=0$ and $g_2 \neq 0$, we rewrite (\ref{eq:determination of uw}) as
\begin{eqnarray}
\frac{\partial v^x}{\partial z} \Big|_{z=0} = \frac{1}{b_2} v^x \Big|_{z=0}
\label{eq:traditional partial slip boundary condition b2}
\end{eqnarray}
with 
\begin{eqnarray}
b_2 \equiv \frac{2\eta}{g_2 v^x \Big|_{z=0}},
\label{eq:b2}
\end{eqnarray}
which implies that we need to treat the macroscopic-velocity-field-dependent slip length. We note that $b_2 \to \infty$ as $v^x |_{z=0} \to \infty$. This divergence stems from the $L$-dependence of $u_w$ given by (\ref{eq:Ldependence of uw L12}).

In summary, when we focus on the $O(L^{-1})$ terms of $\bar{v}(\bar{z})$ in the quasi-equilibrium limit (\ref{eq:infinite volume limit A}), we impose the stick boundary condition  $u_w=0$, the partial slip boundary condition (\ref{eq:traditional partial slip boundary condition b1}), or the perfect slip boundary condition (\ref{eq:traditional partial slip boundary condition b2}) in accordance with the analyticity of $g(u_w)$ at $u_w=0$.

\subsection{Macroscopic boundary condition: hydrodynamic limit}
\label{sec:Macroscopic boundary condition : hydrodynamic limit}
The second macroscopic limit is the hydrodynamic limit:
\begin{eqnarray}
L \to \infty, \ \frac{U}{L} = \const, \ \rho = \const.
\label{eq:infinite volume limit B}
\end{eqnarray}
We focus on the $O(L)$ terms of the velocity fields $\bar{v}^x(\bar{z})$ as the scale of interest. In this section, $\simeq$ indicates equality up to $o(L)$ terms.

We introduce two boundary conditions, stick and perfect slip, in terms of the $L$-dependence of $u_w$; they are defined respectively as
\begin{eqnarray}
u_w = o(L),
\label{eq:def of the stick boundary condition : B},
\end{eqnarray}
and 
\begin{eqnarray}
u_w = O(L)
\label{eq:def of the perfect slip boundary condition : B}
\end{eqnarray}
in the hydrodynamic limit (\ref{eq:infinite volume limit B}). By recalling (\ref{eq:bar vx}), we obtain
\begin{eqnarray}
\bar{v}^x(\bar{z}) \simeq U \bar{z}
\end{eqnarray}
for the stick boundary condition (\ref{eq:def of the stick boundary condition : B}). Therefore, we confirm that the stick boundary condition is consistent with the standard stick boundary condition in hydrodynamics. 

We focus on a relationship between the $L$-dependence of $u_w$ and the functional form of $g(u_w)$. As in the case of Sec.~\ref{sec:Macroscopic boundary condition : quasi-equilibrium limit}, we consider the case in which the functional form of $g(u_w)$ is given by Fig.~\ref{fig:type 1-2 and 1-3}. In Fig.~\ref{fig:type 1-2}, we present the asymptotic behavior of the solution of (\ref{eq:determination of uw}) in the hydrodynamic limit (\ref{eq:infinite volume limit B}), which is in contrast to Fig.~\ref{fig:type 1-3} in the quasi-equilibrium limit (\ref{eq:infinite volume limit A}).
By noting that $\eta U/L$ is fixed and $U$ goes to infinity in the hydrodynamic limit (\ref{eq:infinite volume limit B}), we find that $u_w$ approaches finite value $u_w^{\ast}$ (See Fig.~\ref{fig:type 1-2}). $u_w^{\ast}$ is given by the solution of the equation 
\begin{eqnarray}
\eta \frac{U}{L} = g(u_w^{\ast}).
\label{eq:uwast : hydrodynamic limit}
\end{eqnarray}
When the behavior of $g(u_w)$ is not obtained beyond a linear response regime in $u_w$, it is difficult to determine a concrete value for $u_w^{\ast}$. Nevertheless, we find $u_w^{\ast}$ to be of order $L^0$ from Fig.~\ref{fig:type 1-2}. This corresponds to the stick boundary condition. 

\begin{figure*}[htbp]
\begin{center}
\subfigure[]{
\includegraphics[width=0.48\linewidth]{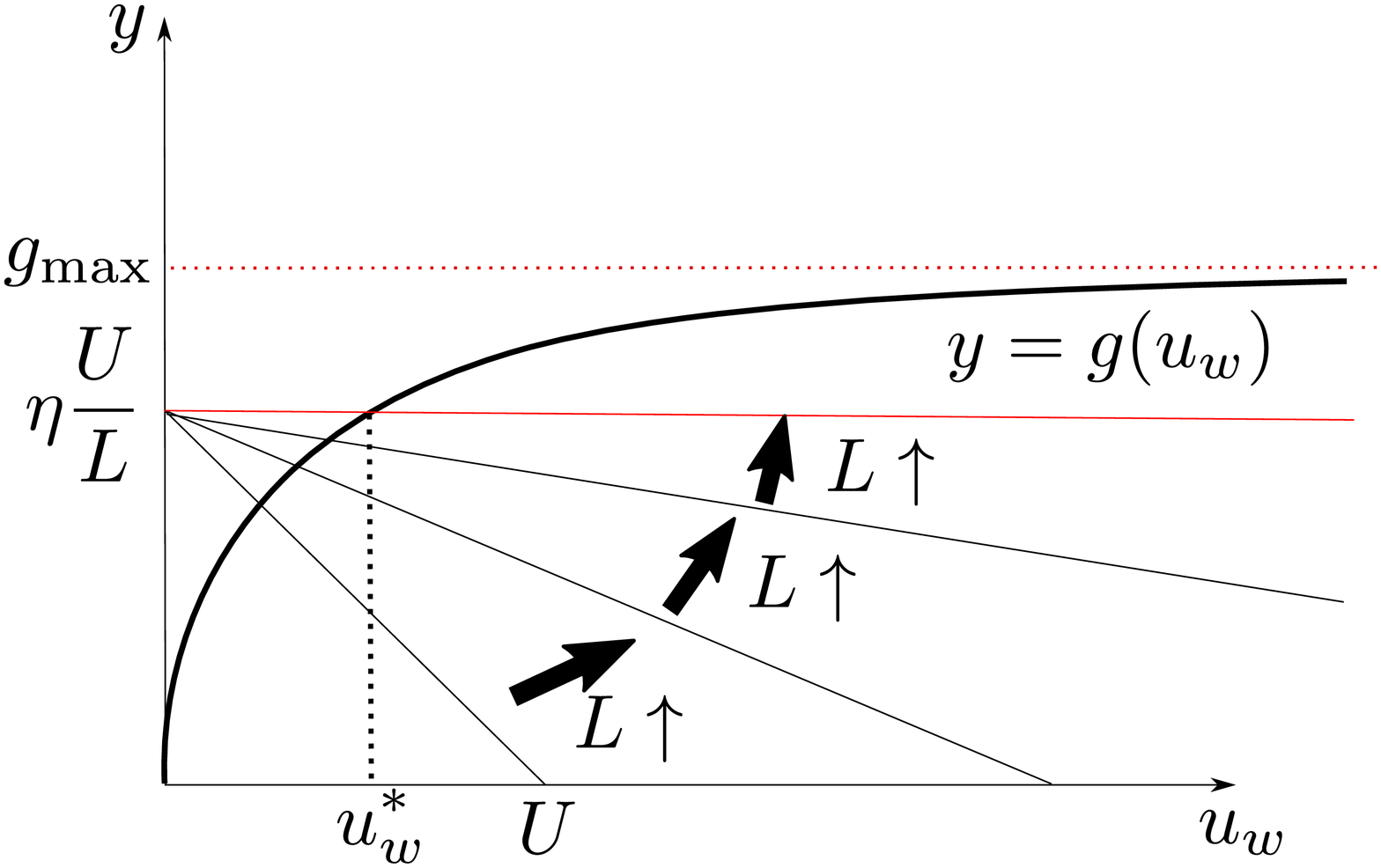} 
\label{fig:type 2-1}}
\subfigure[]{
\includegraphics[width=0.46\linewidth]{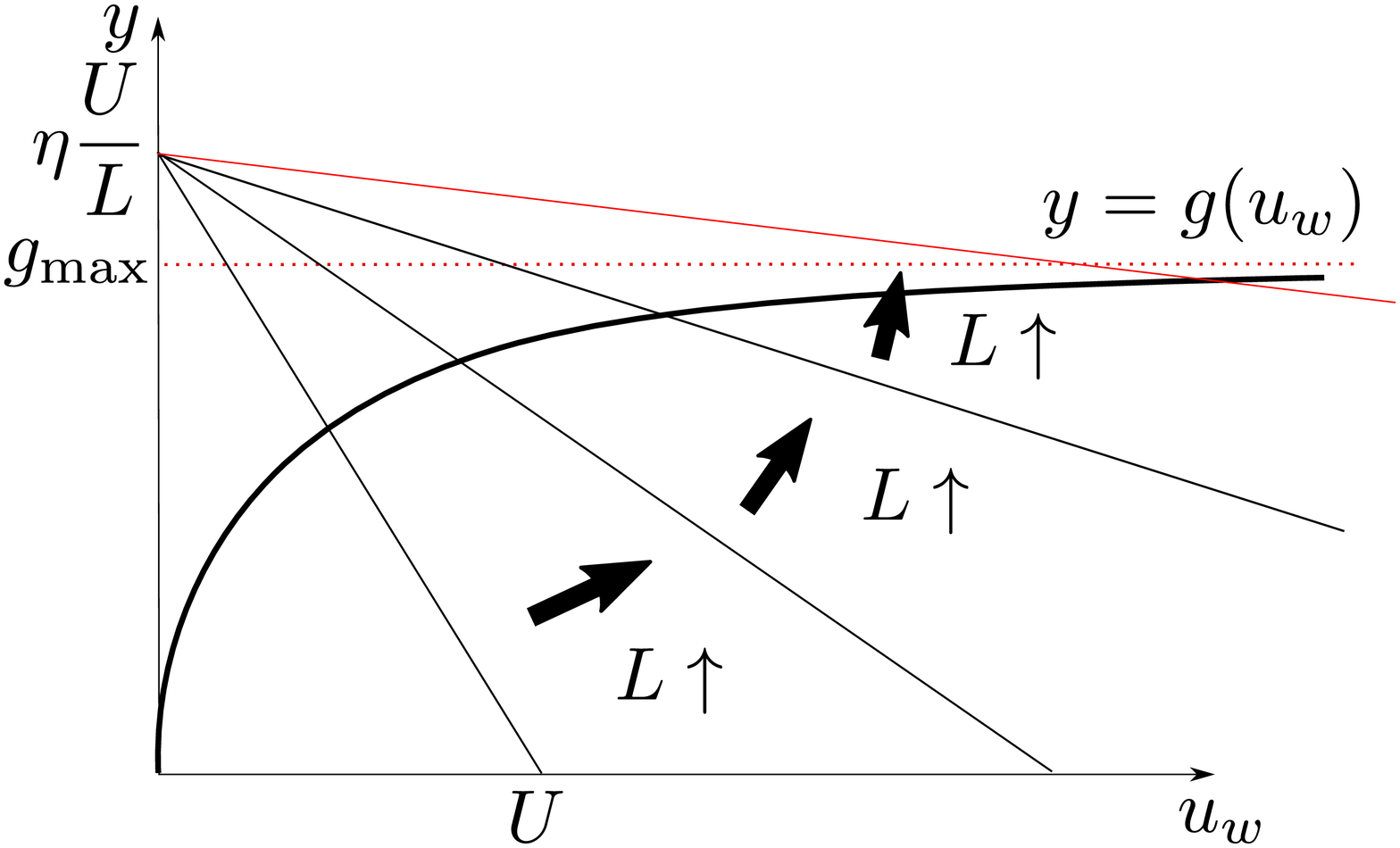}
\label{fig:type 2-2}}
\end{center}
\caption{Schematic graph of $y=g(u_w)$ and $y=\eta (U-u_w)/L$. The intersection of these graphs corresponds to the solution in (\ref{eq:determination of uw}): (a) behavior of the solution for $\eta U/L$ satisfying (\ref{eq:condition: stick boundary condition}) in the hydrodynamic limit; (b) behavior of the solution for $\eta U/L$ satisfying (\ref{eq:condition: perfect slip boundary condition}) in the hydrodynamic limit }
\label{fig:type 2-1 and 2-2}
\end{figure*}

Next, we consider the case in which $g(u_w)$ has a maximum $g_{\max}$. We then find that the $L$-dependence of $u_w$ is classified into two cases depending on $\eta U/L$. We consider the functional form of $g(u_w)$ given by Fig.~\ref{fig:type 2-1 and 2-2}. Fig.~\ref{fig:type 2-1 and 2-2} presents the schematic graph of $y=\eta (U-u_w)/L$ and $y = g(u_w)$. $g(u_w)$ has a maximum value $g_{\max}$ at infinity $u_w \to \infty$. The intersection of the two graphs corresponds to the solution of (\ref{eq:determination of uw}). In Fig.~\ref{fig:type 2-1}, we present the asymptotic behavior of the solution of (\ref{eq:determination of uw}) for
\begin{eqnarray}
\eta \frac{U}{L} < g_{\max}.
\label{eq:condition: stick boundary condition}
\end{eqnarray}
From Fig.~\ref{fig:type 2-1}, we find that $u_w$ approaches finite value $u_w^{\ast}$ independent of $L$, which corresponds to the stick boundary condition. 

In Fig.~\ref{fig:type 2-2}, we present the asymptotic behavior of the solution of (\ref{eq:determination of uw}) for
\begin{eqnarray}
\eta \frac{U}{L} \geq g_{\max}.
\label{eq:condition: perfect slip boundary condition}
\end{eqnarray}
From Fig.~\ref{fig:type 2-2}, we find that $u_w$ goes to infinity in the hydrodynamic limit (\ref{eq:infinite volume limit B}). $O(L)$ terms of $u_w$ are given by
\begin{eqnarray}
u_w \simeq U - L\frac{g_{\max}}{\eta}
\label{eq:u_w in type B perfect slip boundary condition}
\end{eqnarray}
in the hydrodynamic limit (\ref{eq:infinite volume limit B}), which corresponds to the perfect slip boundary condition. Note that (\ref{eq:u_w in type B perfect slip boundary condition}) is rewritten in terms of the shear stress as
\begin{eqnarray}
\sigma^{xz} \simeq g_{\max}.
\label{eq:stress expression in type B perfect slip boundary condition}
\end{eqnarray}
When $g_{\max}=0$, (\ref{eq:stress expression in type B perfect slip boundary condition}) corresponds to the standard perfect slip boundary condition imposed on the solutions of the Euler equation.
These results indicate that the boundary condition depends on the behavior of $g(u_w)$ over the entire range, i.e., the existence of the maximum.

\begin{figure}[h]
\begin{center}
\includegraphics[width=0.7\linewidth]{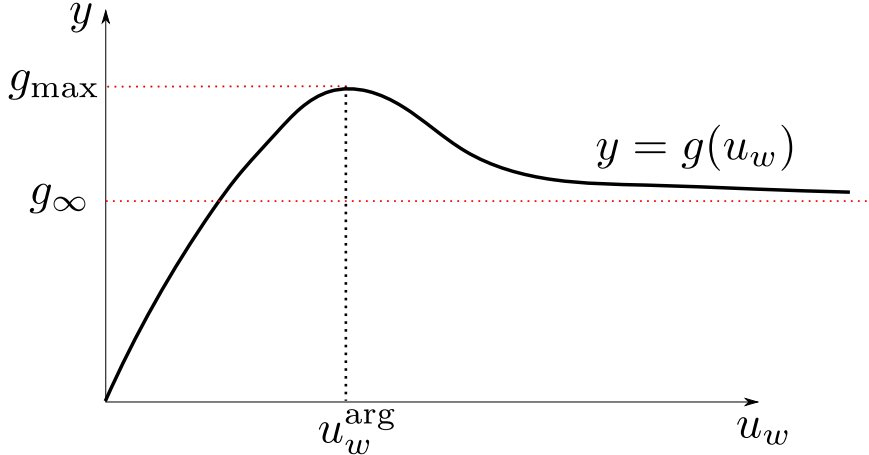}
\end{center}
\caption{Schematic image of $g(u_w)$ for the wall $a=0.5$, $c_{\rm FS}=0.6$.}
\label{fig:type 3}
\end{figure}

Finally, we consider the case that $g(u_w)$ is given by Fig.~\ref{fig:uw vs gamma}. We conjecture that the functional form of $g(u_w)$ in Fig.~\ref{fig:uw vs gamma} is given by Fig.~\ref{fig:type 3}. That is, $g(u_w)$ has a maximum value $g_{\max}$ at $u_w=u_w^{\arg}$ and approaches a constant value $g_{\infty} \leq g_{\max}$ as $u_w \to \infty$.
By a similar procedure to that in Fig.~\ref{fig:type 2-1 and 2-2}, we find that there are three solutions of (\ref{eq:determination of uw}) depending on $\eta U/L$. First, when the following inequality is satisfied
\begin{eqnarray}
\eta \frac{U}{L} < g_{\infty},
\end{eqnarray}
the solution of (\ref{eq:determination of uw}) approaches finite value $u_w^{\ast}$ in the hydrodynamic limit (\ref{eq:infinite volume limit B}), which is given by (\ref{eq:uwast : hydrodynamic limit}). This implies that the stick boundary condition applies. Second, when
\begin{eqnarray}
g_{\infty} \leq \eta \frac{U}{L} \leq g_{\max}
\end{eqnarray}
holds, (\ref{eq:determination of uw}) has three solutions in the hydrodynamic limit (\ref{eq:infinite volume limit B}). In the hydrodynamic limit (\ref{eq:infinite volume limit B}), the two smaller solutions approach finite values whereas the largest solution diverges. We consider that all three solutions are physically realizable. In particular, we anticipate that when the external force increases sufficiently slowly from $0$ to the appropriate value, the smallest solution is realized. Since the smallest solution approaches a finite value in the hydrodynamic limit (\ref{eq:infinite volume limit B}), this solution corresponds to the stick boundary condition. Finally, when
\begin{eqnarray}
g_{\max} < \eta \frac{U}{L}
\end{eqnarray}
holds, (\ref{eq:determination of uw}) has one solution, which goes to infinity in the hydrodynamic limit (\ref{eq:infinite volume limit B}). This corresponds to the perfect slip boundary condition, which, in terms of shear stress, is written
\begin{eqnarray}
\sigma^{xz} \simeq g_{\infty}.
\label{eq:stress expression in type C perfect slip boundary condition}
\end{eqnarray}

In summary, when we focus on the $O(L)$ terms of $\bar{v}(\bar{z})$ in the hydrodynamic limit (\ref{eq:infinite volume limit B}), we impose either the stick boundary condition  $u_w=0$ or the perfect slip boundary condition (\ref{eq:stress expression in type B perfect slip boundary condition}) (or (\ref{eq:stress expression in type C perfect slip boundary condition})) in accordance with the behavior of $g(u_w)$ over its entire range.

\section{summary}
\label{sec:summary}
In this paper, we proposed the boundary conditions appropriate for macroscopic hydrodynamics. The key idea of our study was to separate the microscopic boundary condition, which is uniquely determined from the microscopic description of the fluid and the wall, and the macroscopic boundary condition, which depends on the scale of interest. We studied the macroscopic boundary conditions based on the microscopic boundary condition and the macroscopic limits for non-equilibrium steady states.

We used (\ref{eq:constitutive equation in boundary region}) as the microscopic boundary condition, because (\ref{eq:constitutive equation in boundary region}) is the simplest boundary condition satisfying locality. Here, $g(u_w)$ is uniquely determined from the microscopic parameters of the fluid and the wall. We showed that $g(u_w)$ has maximum value for our model using the molecular dynamical simulation.

With ignoring higher terms of $\bar{v}^x(\bar{z})$ in $L$, we introduced the scaled velocity field that depends on the scale of interest. The macroscopic boundary condition is determined so that the standard fluid dynamics with it gives the scaled velocity field. We proposed two frameworks for determining the macroscopic boundary conditions by defining two macroscopic limits.

The first macroscopic limit is the quasi-equilibrium limit. By focusing on the $O(L^{-1})$ terms of the velocity fields $\bar{v}^x(\bar{z})$, we constructed a framework to describe the macroscopic boundary condition comprising three boundary conditions: stick, partial slip, and perfect slip. We showed that the boundary conditions are determined only by the analyticity of $g(u_w)$ at $u_w=0$. Then, we may classify the boundary conditions in terms of the $u_w$-dependence of the slip length. The stick boundary condition corresponds to $b=0$. The partial slip boundary condition corresponds to the $u_w$-independent finite slip length: (\ref{eq:traditional partial slip boundary condition b1}) with (\ref{eq:b1}). The perfect slip boundary condition corresponds to the $u_w$-dependent slip length: (\ref{eq:traditional partial slip boundary condition b2}) with (\ref{eq:b2}).

The second macroscopic limit is the hydrodynamic limit. By focusing on the $O(L)$ terms of the velocity fields $\bar{v}^x(\bar{z})$, we established a framework for the macroscopic boundary condition that contains two boundary conditions: stick and perfect slip. We showed that the boundary conditions are related to the behavior of $g(u_w)$ over the entire range such as $g_{\max}$ and $g_{\infty}$. We applied this framework to three cases with $g(u_w)$ of the form given by Figs.~\ref{fig:type 1-2 and 1-3}, \ref{fig:type 2-1 and 2-2} and \ref{fig:type 3}. When $g(u_w)$ is given by Fig.~\ref{fig:type 2-1 and 2-2}, the stick boundary condition $u_w=0$ is realized in the case $\eta U/L < g_{\max}$, whereas the perfect slip boundary condition $\sigma^{xz}=g_{\max}$ is realized in the case $\eta U/L > g_{\max}$.

\section{discussion}
\label{sec:discussion}
Let us remark on the macroscopic boundary condition for systems with more general geometries in the hydrodynamic limit. The result in Sec.~\ref{sec:Macroscopic boundary condition : hydrodynamic limit} contains the configuration-dependent quantity $\eta U/L$. We conjecture that, by replacing $U/L$ with the shear rate assuming the stick boundary condition, the discussion in Sec.~\ref{sec:Macroscopic boundary condition : hydrodynamic limit} also applies to more general configurations. Based on this conjecture, we obtain the framework in the hydrodynamic limit when $g(u_w)$ is given by Fig.~\ref{fig:type 2-1 and 2-2}. Specifically, we start by assuming the stick boundary condition:
\begin{eqnarray}
\bm{v}\cdot \bm{\tau}\Big|_{s} = 0,
\end{eqnarray}
where $\bm{\tau}$ is the tangential vector of the surface and the subscript $s$ represents the evaluation at the surface. When
\begin{eqnarray}
\sigma^{ij}\tau_i n_j \Big|_{s} \geq g_{\max}
\end{eqnarray}
holds, where the left-hand side is calculated on the stick boundary condition and $\bm{n}$ is the normal vector of the surface, we apply the perfect boundary condition
\begin{eqnarray}
\sigma^{ij}\tau_i n_j \Big|_{s} = g_{\max}.
\label{eq:perfect slip boundary condition}
\end{eqnarray}

Our concept of the macroscopic boundary condition may be applied to laboratory experiments. Recently, the slip phenomena were confirmed to be important for nano- and micro- scale systems~\cite{neto2005boundary,lauga2007microfluidics,cao2009molecular,bocquet2010nanofluidics}. One of the reasons why the slip length is regarded as an important quantity in small systems is that the observations are done with the high accuracy for such systems. We consider that the framework of the quasi-equilibrium limit is useful to explain phenomena in such small systems, because we can calculate $O(L^{-1})$ terms of $u_w/U$ by using one given parameter $g_1$ as shown in (\ref{eq:uw partial slip}). This is in contrast to the framework established under the hydrodynamic limit, which requires more information about $g(u_w)$ to calculate $O(L^{-1})$ terms of $u_w/U$ as shown in (\ref{eq:uwast : hydrodynamic limit}). As $L$ is increased with $U$ fixed and observations are done with lower accuracy, we may ignore even $O(L^{-1})$ terms of the velocity fields $\bar{v}^x(\bar{z})$ in the quasi-equilibrium limit. However, when $U$ is sufficiently large, we consider that the slip phenomena are important even for such large systems. In general, when we apply a high shear stress to a fluid, we may observe the slip length of the order of micrometers with the non-linearity~\cite{choi2003apparent,ulmanella2008molecular,thompson1997general,priezjev2004molecular,martini2008slip,priezjev2009shear,priezjev2010relationship}. In such situations, we consider it useful to apply the framework established under the hydrodynamic limit, because it is the simplest framework to extract non-linear behavior of $g(u_w)$.

As a related study, Priezjev {\it et al.} reported the shear-rate-dependence of slip length in the shear flow of polymer melts past atomically smooth surfaces~\cite{priezjev2004molecular,priezjev2009shear,priezjev2010relationship}. By using the molecular dynamical simulation, they demonstrated that $g_{\max}$ decreases with increasing the chain length and  is nearly independent of the chain length beyond ten bead-spring units~\cite{priezjev2004molecular}. It was also found that the onset of the non-linear regime of polymer melts is observed at lower shear rates than that of simple liquids~\cite{priezjev2010relationship}. We expect that the macroscopic boundary condition is useful at high shear rates in these systems. In order to realize a macroscopic slip in realistic systems beyond small systems in a laboratory, it is important to quantitatively evaluate $g_{\rm max}$, $g_{\infty}$ and $u_w^{\rm arg}$ of various type of fluid under realistic settings.

Particularly, for the dilute gases, the slip phenomena have been studied theoretically and experimentally. It was found that when the Knudsen number is on the order of $0.001$ or larger, non-negligible slip occurs~\cite{colin2005,maali2016slip}. Recent experiments reported the slip length of $500{\rm nm}$~\cite{seo2013situ}. The microscopic boundary condition for the gas flow has been discussed by numerous researchers and various slip boundary conditions have been proposed in the literature~\cite{colin2005}. They are more complicated than the microscopic boundary condition (\ref{eq:constitutive equation in boundary region}) assumed in this paper. Therefore, it is difficult to apply the results obtained in this paper to the gas flow. However, we consider that the idea to introduce the macroscopic boundary conditions is still useful, because the relatively complicated boundary conditions are expressed as simpler boundary conditions with a few parameters characterizing an amount of slip and non-linearity. Developing the macroscopic boundary condition for the gas flow is the next problem.

\section*{acknowledgment}
The authors thank A. Yoshimori and Y. Minami for helpful comments. The present study was supported by JSPS KAKENHI Grant Numbers JP17H01148.

\appendix
\section{Expression of microscopic density fields and microscopic currents}
\label{sec:expression of microscopic density}
The microscopic mass density field $\hat{\rho}(\bm{r};\Gamma)$ and the microscopic momentum density field $\hat{\pi}^a(\bm{r};\Gamma)$ are defined as
\begin{eqnarray}
\hat{\rho}(\bm{r};\Gamma) \equiv \sum_{i} m \delta(\bm{r}-\bm{r}_i),
\end{eqnarray}
\begin{eqnarray}
\hat{\pi}^a(\bm{r};\Gamma) \equiv \sum_{i} p_i^a \delta(\bm{r}-\bm{r}_i).
\end{eqnarray}
We assume that the $z=0$ wall consists of $N_w$ material points and $U_{BW}(\bm{r}_i)$ is given by (\ref{eq:UBW}). Then, $\hat{\pi}^a(\bm{r},\Gamma)$ satisfies the continuity equation\cite{spohn2012large,das2011statistical,sasa2014derivation}
\begin{eqnarray}
\frac{\partial \hat{\pi}^a(\bm{r};\Gamma_t)}{\partial t} + \frac{\partial \hat{J}^{ab}(\bm{r};\Gamma_t)}{\partial r^b} = 0
\label{eq:continue equation}
\end{eqnarray}
in $0<z<L$, where the microscopic momentum current $\hat{J}^{ab}(\bm{r};\Gamma)$ is given by
\begin{eqnarray}
\hat{J}^{ab}(\bm{r};\Gamma) \equiv \hat{J}^{ab}_b(\bm{r};\Gamma) + \hat{J}^{ab}_w(\bm{r};\Gamma)
\end{eqnarray}
with
\begin{eqnarray}
\hat{J}^{ab}_b(\bm{r};\Gamma) &\equiv& \sum_{i} \frac{p^a_i p^b_i}{m} \delta(\bm{r}-\bm{r}_i) + \sum_{i<j} F_{ij}^a(r^b_i-r^b_j) D(\bm{r};\bm{r}_i,\bm{r}_j),
\end{eqnarray}
\begin{eqnarray}
\hat{J}^{ab}_w(\bm{r};\Gamma) &\equiv& \sum_{i=1}^N \sum_{j=1}^{N_w} F_{ij}^{w a}(r^b_i-q^b_j) D(\bm{r};\bm{r}_i,\bm{q}_j),
\end{eqnarray}
where we have used the definition of the following quantities:
\begin{eqnarray}
D(\bm{r};\bm{r}_i,\bm{r}_j) \equiv \int_0^1 d \xi \delta(\bm{r}-\bm{r}_i-(\bm{r}_j-\bm{r}_i)\xi),
\end{eqnarray}
\begin{eqnarray}
F_{ij}^a \equiv -\frac{\partial V_{\rm FF}(|\bm{r}_i-\bm{r}_j|)}{\partial r^a_i},
\end{eqnarray}
\begin{eqnarray}
F_{ij}^{wa} \equiv -\frac{\partial V_{\rm BW}(|\bm{r}_i-\bm{q}_j|)}{\partial r^a_i}.
\end{eqnarray}

\begin{figure}[h]
\begin{center}
\includegraphics[width=0.6\linewidth]{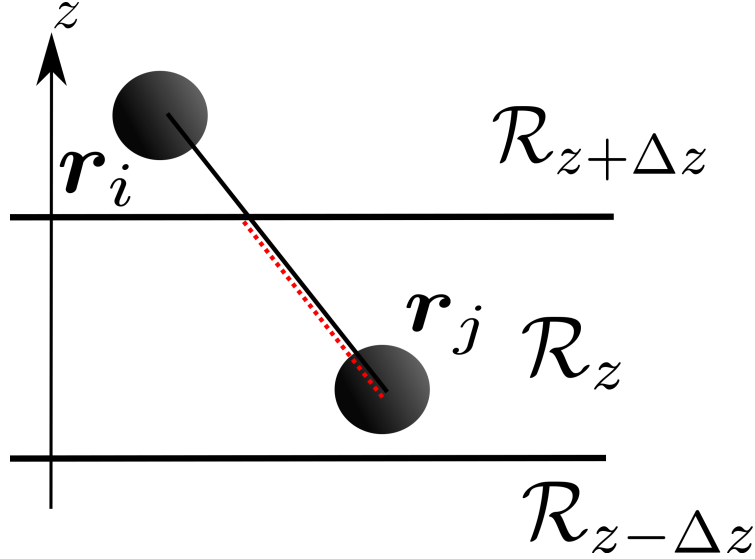}
\end{center}
\caption{Schematic image of $D(\bm{r}_i,\bm{r}_j;\mathcal{R}_z)$. $D(\bm{r}_i,\bm{r}_j;\mathcal{R}_z)$ is given as the ratio of black line and red dotted line. }
\label{fig:DefofD}
\end{figure}

In the numerical simulation, the averaged density fields are calculated by spatially and temporally averaging the microscopic density fields (e.g., (\ref{eq:example of rho})). These quantities are expressed as
\begin{eqnarray}
\rho(z) = \frac{1}{\tau}\int_0^{\tau} dt \frac{1}{L_xL_y \Delta z} \sum_{i} m\Theta(\bm{r}_i(t);\mathcal{R}_z),
\end{eqnarray}
\begin{eqnarray}
\pi^a(z) = \frac{1}{\tau}\int_0^{\tau} dt \frac{1}{L_xL_y \Delta z} \sum_{i} p_i^a\Theta(\bm{r}_i(t);\mathcal{R}_z),
\end{eqnarray}
\begin{eqnarray}
& & J^{ab}_b(z) = \frac{1}{\tau}\int_0^{\tau} dt \frac{1}{L_xL_y \Delta z} \Big[ \sum_{i} \frac{p^a_i(t) p^b_i(t)}{m} \Theta(\bm{r}_i(t);\mathcal{R}_z) \nonumber \\[3pt]
& & \hspace{+5cm} + \sum_{i<j} F_{ij}^a(t)(r^b_i(t)-r^b_j(t)) D(\bm{r}_i(t),\bm{r}_j(t);\mathcal{R}_z) \Big],
\end{eqnarray}
and
\begin{eqnarray}
J^{ab}_w(z) \nonumber = \frac{1}{\tau}\int_0^{\tau} dt \frac{1}{L_xL_y \Delta z} \Big[ \sum_{i=1}^N \sum_{j=1}^{N_w}  F_{ij}^{wa}(t)(r^b_i(t)-r^b_j(t)) D(\bm{r}_i(t),\bm{r}_j(t);\mathcal{R}_z) \Big]
\end{eqnarray}
with
\begin{eqnarray}
\mathcal{R}_z = [0,L_x]\times[0,L_y]\times[z-\frac{\Delta z}{2},z+\frac{\Delta z}{2}],
\end{eqnarray}
\begin{eqnarray}
\Theta(\bm{r};\mathcal{R}_z) = \begin{cases}
0 & \bm{r} \not\in \mathcal{R}_z, \\
1 & \bm{r} \in \mathcal{R}_z ,
\end{cases}
\end{eqnarray}
and
\begin{eqnarray}
D(\bm{r}_i,\bm{r}_j;\mathcal{R}_z) = \begin{cases}
0 & \bm{r}_i \in \mathcal{R}_{z-\Delta z} \ {\rm and} \  \bm{r}_j \in \mathcal{R}_{z-\Delta z} ,\\
z_{jd}/z_{ij} & \bm{r}_i \in \mathcal{R}_{z-\Delta z} \ {\rm and} \  \bm{r}_j \in \mathcal{R}_{z}, \\
(z_{ij}-z_{id}-z_{ju})/z_{ij} & \bm{r}_i \in \mathcal{R}_{z-\Delta z} \ {\rm and} \  \bm{r}_j \in \mathcal{R}_{z+\Delta z}, \\
z_{id}/z_{ij} & \bm{r}_i \ \in \mathcal{R}_{z} {\rm and} \  \bm{r}_j \in \mathcal{R}_{z-\Delta z}, \\
1 & \bm{r}_i \in \mathcal{R}_z \ {\rm and} \  \bm{r}_j \in \mathcal{R}_z, \\
z_{iu}/z_{ij} & \bm{r}_i \ \in \mathcal{R}_{z} {\rm and} \  \bm{r}_j \in \mathcal{R}_{z+\Delta z}, \\
(z_{ij}-z_{jd}-z_{iu})/z_{ij} & \bm{r}_i \ \in \mathcal{R}_{z+\Delta z} {\rm and} \  \bm{r}_j \in \mathcal{R}_{z-\Delta z}, \\
z_{ju}/z_{ij} & \bm{r}_i \ \in \mathcal{R}_{z+\Delta z} {\rm and} \  \bm{r}_j \in \mathcal{R}_{z} ,\\
0 & \bm{r}_i \ \in \mathcal{R}_{z+\Delta z} {\rm and} \  \bm{r}_j \in \mathcal{R}_{z+\Delta z}, \\
\end{cases}
\end{eqnarray}
where $r_{ij} = |\bm{r}_i-\bm{r}_j|$, $z_{ij} = |z_i-z_j|$, $z_{id} = |z_i-(z-\Delta z/2)|$ and $z_{iu} = |z_i-(z+\Delta z/2)|$. Here, we give the graphical interpretation of $D(\bm{r}_i,\bm{r}_j;\mathcal{R}_z)$ in Fig.~\ref{fig:DefofD}.

\section{Scaling law at high shear rates}
\label{sec:scaling law at high shear rates}
In Sec.~\ref{sec:Molecular dynamical simulation}, we gave an example for which $g(u_w)$ has a maximum (see Fig.~\ref{fig:uw vs gamma}). We conjecture that the functional form of $g(u_w)$ is given by Fig.~\ref{fig:type 3}. With this conjecture, we considered the macroscopic boundary condition in Sec.~\ref{sec:Macroscopic boundary condition : hydrodynamic limit}. As explained in Sec.~\ref{sec:Microscopic boundary condition}, the previous studies~\cite{thompson1997general,priezjev2004molecular,priezjev2009shear} reported the behavior that $b(\dot{\gamma}_w)$ diverges at $\dot{\gamma}_w \to \dot{\gamma}_c$ and provided the scaling law for simple liquids
\begin{eqnarray}
\frac{b(\dot{\gamma}_w)}{b^{\ast}} = \Big(1 - \frac{\dot{\gamma_w}}{\dot{\gamma}_c} \Big)^{-\frac{1}{2}}
\label{eq:scaling relation}
\end{eqnarray}
near the critical value $\dot{\gamma}_c$, where $b^{\ast}$ is a constant. In this Appendix, we show that, under some assumptions, $g(u_w)$ of the type shown in Fig.~\ref{fig:type 3} satisfies the scaling law (\ref{eq:scaling relation}). That is, our simulations are consistent with the previous studies in terms of scaling behavior.

As $U$ increases from $0$, $u_w$ increases and $g$ goes up a slope of $g(u_w)$ to reach $g_{\max}$ (See Fig.~\ref{fig:type 3}). We assume that $u_{w}^{\arg}$ is sufficiently large so that $u_w$ cannot reach $u_w^{\arg}$ within numerical simulations. Then, we restrict ourselves to $u_w < u_w^{\arg}$. Since $g(u_w)$ is a bijective function in $u_w < u_w^{\arg}$, we rewrite (\ref{eq:h g inverse}) in terms of the inverse function of $g(u_w)$ as
\begin{eqnarray}
g^{-1}(\eta \dot{\gamma}_w) = b(\dot{\gamma}_w) \dot{\gamma}_w.
\label{eq:ginverse b}
\end{eqnarray}
We introduce the critical shear rate $\dot{\gamma}_c$ by
\begin{eqnarray}
\dot{\gamma}_c = \frac{g_{\max}}{\eta} .
\label{eq:gc gmax}
\end{eqnarray}
By noting that $g^{-1}(\eta \dot{\gamma}_c) =u_w^{\arg}$, we obtain
\begin{eqnarray}
\frac{d g^{-1}(\eta \dot{\gamma}_w)}{d \dot{\gamma}_w} \Big|_{\dot{\gamma}_c} = \Big(\frac{d g(u_w)}{du_w} \Big|_{u_w^{\arg}} \Big)^{-1} = \infty.
\label{eq:derivative ginverse}
\end{eqnarray}
From (\ref{eq:ginverse b}) and (\ref{eq:derivative ginverse}), we find that $\dot{\gamma}_c$ is the point for which the first derivative of $b(\dot{\gamma}_w)$ in $\dot{\gamma}_w$, $db(\dot{\gamma}_w)/d \dot{\gamma}_w$ diverges.

If we regard $u_w^{\arg}$ as infinity, then we conjecture that $g(u_w)$ can be expanded around $u_w=u_w^{\arg}$ as
\begin{eqnarray}
g(u_w) = g_{\max} - \Big( \frac{1}{u^4_w} \frac{d^2 g}{d u_w^2} \Big)_{u_w^{\arg}} \frac{1}{u_w^2} + \cdots .
\label{eq:expansion gamma infinite}
\end{eqnarray}
Equation (\ref{eq:expansion gamma infinite}) means that $g(u_w)$ has no singular point near $u_w = u_w^{\arg}$. By using (\ref{eq:ginverse b}), (\ref{eq:gc gmax}), and (\ref{eq:expansion gamma infinite}), we find that $b(\dot{\gamma}_w)$ is given by
\begin{eqnarray}
b(\dot{\gamma}_w) \simeq \frac{1}{\dot{\gamma}_c\sqrt{g_{\max}}} \sqrt{\Big( \frac{1}{u^4_w} \frac{d^2 g}{d u_w^2} \Big)_{u_w^{\arg}}} \Big(1- \frac{\dot{\gamma}_w}{\dot{\gamma}_c} \Big)^{-\frac{1}{2}}
\label{eq:B6}
\end{eqnarray}
near $\dot{\gamma}_w=\dot{\gamma}_c$. Equation (\ref{eq:B6}) implies that $b(\dot{\gamma}_w)$ diverges at $\dot{\gamma}_w=\dot{\gamma}_c$ following the scaling law (\ref{eq:scaling relation}). Thus, we conclude that the results of our simulation are consistent with the previous studies in terms of scaling behavior.

\bibliography{EV11690}

\end{document}